\documentclass[journal]{IEEEtran}

\usepackage{bibentry}  
\usepackage{xcolor,soul,framed} 

\colorlet{shadecolor}{yellow}
\usepackage[pdftex]{graphicx}
\graphicspath{{../pdf/}{../jpeg/}}
\DeclareGraphicsExtensions{.pdf,.jpeg,.png}

\usepackage[cmex10]{amsmath}
\usepackage{subcaption}
\usepackage{array}
\usepackage{mdwmath}
\usepackage{mdwtab}
\usepackage{eqparbox}
\usepackage{url}
\usepackage[T1]{fontenc} 
\usepackage{amsmath}
\usepackage[cmintegrals]{newtxmath}
\usepackage{bm} 
\usepackage{comment}
\RequirePackage{algorithm2e}

\begin{document}

\bstctlcite{IEEEexample:BSTcontrol}
    \title{Terminal Phase Navigation for AUV Docking: An Innovative Electromagnetic Approach}
  \author{Yevgeni~Gutnik and Morel~Groper.\\
  yevgenigutnik@gmail.com\\
      The Hatter Department of Marine Technologies, Leon H. Charney School of Marine Sciences, University of Haifa, Israel}


  
\maketitle
\begin{abstract}

This study introduces a groundbreaking approach for real-time 3D localization, specifically focusing on achieving seamless and precise localization during an AUV's terminal guidance phase as it approaches an omnidirectional docking component in an automated Launch and Recovery System (LARS). Through the use of the AUV's magnetometer, an economical electromagnetic beacon embedded in the docking station, and an advanced signal processing algorithm, this novel approach ensures the accurate localization of the docking component in three dimensions without the need for direct line-of-sight contact. The method's real-time capabilities were rigorously evaluated via simulations, prototype experiments in a controlled lab setting, and extensive full-scale pool experiments. These assessments consistently demonstrated an exceptional average positioning accuracy of under 3 cm., marking a significant advancement in AUV guidance systems.

\end{abstract}


%
\IEEEpeerreviewmaketitle

\section{Introduction}
\IEEEPARstart Autonomous underwater vehicles (AUVs) are unmanned, untethered, self-propelled, and self-controlled robots. These vehicles are capable of operating independently for extended periods without the need for continuous manned supervision, while efficiently collecting data from a wide variety of sensors, making them highly valuable for scientific, commercial, and military applications. 

However, due to their untethered operation, the endurance of AUVs is inherently constrained by the capacity of their onboard batteries and data storage systems. Consequently, periodic recovery operations are necessary to facilitate the recharging of power sources and the transfer of stored data. Traditionally, these launch and recovery (L\&R) procedures have been performed with human intervention, involving the latching and lifting of AUVs from the sea surface. Surface L\&R operations, particularly in adverse weather conditions and rough seas, pose substantial risks to personnel and equipment.

Autonomous subsurface docking of AUVs can significantly enhance the robustness of L\&R operations by executing a crucial phase of the operation at depths minimally affected by surface waves and wind. This capability eliminates the need to conduct L\&R operations in adverse environmental conditions, thereby extending the operational range of AUVs.

Various approaches to the subsurface docking process have been explored, including the use of protective frames \cite{gu2018automated}, \cite{sarda2016usv}, \cite{jalving2008payload}, \cite{lin2022docking}, flexible wires, rigid poles, and other capture mechanisms \cite{kimball2018artemis}, \cite{sarda2018launch}, \cite{piskura2016development}. Protective frames, while offering enhanced protection, introduce complexity and require precise alignment between the AUV and the docking mechanism. In contrast, flexible wires or rigid poles enable omnidirectional docking but necessitate the installation of dedicated line-capturing devices on the AUV and the use of precise positioning methods for successful docking.

The localization and docking procedure is typically supported by various sensors, including acoustic sensors, vision cameras, electromagnetic (EM) field sensing, or a fusion of multiple sensors \cite{hildebrandt2017combining}. Acoustic sensors offer long-range detection capabilities but have relatively low resolution and update rates ($\approx 1 Hz$) \cite{miranda2013homing}. Vision cameras provide high-precision positioning information but are susceptible to water turbidity and lighting conditions, relying on a continuous line of sight to maintain localization \cite{maki2013docking}, \cite{fan2019auv}.

EM guidance, on the other hand, is immune to water conditions and line-of-sight limitations, making it an ideal choice for the critical final stage of docking known as the terminal guidance phase \cite{kusche2021indoor}, \cite{dai20176}, \cite{pasku2017magnetic}.

\begin{figure}[h]\centering
 \includegraphics[width=\linewidth]{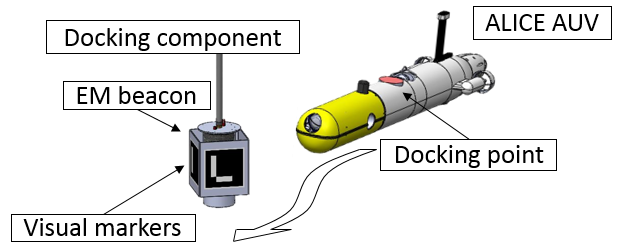}
  \caption{ALICE AUV and the docking component of our LARS system}
  \label{alice_dock}
\end{figure}

Building upon the work presented in \cite{gutnik2023data}, this research introduces a comprehensive framework that integrates the practical implementation of the proposed method to create a highly precise electromagnetic (EM)-based positioning system. This system is specifically designed to achieve accurate 3D positioning during the terminal guidance phase of an AUV docking process. Its primary objective is to guide our ALICE AUV towards an omnidirectional docking component of an automated Launch and Recovery System (LARS), as illustrated in Figure \ref{alice_dock}. 

The proposed method employs an EM beacon integrated into the docking component of the LARS while the ALICE AUV utilizes an onboard magnetometer to measure the magnetic field. Customized signal extraction and positioning computation algorithms were developed to extract the beacon's signals and determine the AUV's 3D position relative to the beacon. Additionally, visual markers on the docking component and an onboard camera enable a unique initialization process, effectively resolving inherent magnetic field computational ambiguities. As a result, this method offers a distinct positioning solution without restricting the AUV's operation to a specific sector of the beacon and without requiring continuous synchronization between the transmitting coils and the receiving magnetometer.

\section{Related works}
EM-based positioning and guidance systems primarily function by detecting an artificial magnetic field generated by dedicated electromagnetic beacons strategically positioned within the operational area, as described by Bian et al. \cite{bian2021induced}. 

Sheinker et al. \cite{sheinker2013localization} developed a 3D EM-based localization method using three magnetic beacons and a tri-axial search coil magnetometer. These beacons emitted modulated signals at specific frequencies. A set of lock-in amplifiers (LIAs) was employed to extract these signals from the magnetometer's sampled magnetic flux field. The study evaluated three different beacon placement configurations through simulations and field tests, achieving a localization error below 0.77 m with a mean error of 0.25 m in controlled field experiments covering an area of 10 m × 11 m.

Andria et al. \cite{andria2019development} developed a 3D EM tracking system for surgical navigation that utilizes five coils generating AC magnetic fields at distinct frequencies and a 6-DOF magnetic probe. Position estimation is achieved through interpolation and fitting of the measured field to calibration points. This approach yielded a maximum mean error of 3.7 mm within a one-meter range from the beacon.

Hu et al. \cite{hu2012novel} introduced an EM guidance system designed for positioning medical instruments during spinal surgeries. Their system employed tri-axial transmitting coils and tri-axial receiving coils which were excited by alternating current at distinct frequencies. A function fitting and optimization approach calculated the 3D position of the beacon with an accuracy of 1-2 mm and an orientation accuracy under $1^\circ$ within a distance of 0.5 meters from the beacon. It is noteworthy, however, that the system's operational range was limited, and complete beacon state determination necessitated prior information. 

Prior studies on EM-based guidance for underwater docking stations have primarily concentrated on the development of two-dimensional systems tailored for directional docking with seabed-fixed stations.
 
Feezor et al. \cite{feezor2001autonomous} presented an EM-based guidance system for directing an AUV into a cone-shaped docking station. The system utilized three transmitting coils placed on the dock and triaxial receiving coils onboard the AUV. Two of the coils were utilized to generate a signal distinguishing the entrance from the rear side of the dock, while the third coil determined the relative bearing between the AUV and the dock's centerline. Successful docking within a range of 25-30 m was achieved; however, it is important to note that this system provided only relative bearing information without offering range data.

Peng et al. \cite{peng2019low} introduced a system featuring a transmitting coil situated on the docking station and triaxial search coils integrated into the AUV. This system was specifically designed for compact and cost-effective AUVs. Utilizing a dedicated signal processing module, the system extracted the amplitude and phase of the transmitting coil, enabling the computation of the bearing between the AUV and the dock entrance with a detection range of 20 m. It's worth noting, however, that the system provided only relative bearing information.

Vandavasi et al. \cite{vandavasi2018concept} introduced a guidance system utilizing a single transmitting coil positioned on a funnel-shaped dock and two magnetometers installed on a small AUV. The determination of the relative bearing between the AUV and the dock entrance relied on analyzing the difference between the two magnetometer measurements. Estimating the range to the dock involved fitting the measured magnetic field to a spatial magnetic map of the transmitting coil, previously computed using the finite element analysis (FEA) method. Successful demonstrations were achieved within an effective range of 7 m. It is important to note, however, that this method required the use of two magnetometers and the prior development of the magnetic field map.

Lin et al. \cite{lin2022underwater} employed a single transmitting coil and triaxial receiving coils to guide an AUV into a funnel-shaped dock. The measured amplitudes and phases were used for calculating both the bearing and the range to the dock. To address intensity ambiguity, the guidance was confined to the horizontal plane within an angular range of approximately $10^\circ$ relative to the dock's center line. Successful docking scenarios were achieved with an accuracy of less than 0.2 m and orientation within $2.5^\circ$ within a 6-meter range.

The subsequent sections of this paper are organized as follows: In Section~\ref{Mathematical_Model}, a detailed overview of the physical structure and mathematical background of the proposed EM-based localization and guidance system is provided. Section~\ref{signal_processing} outlines the signal extraction and processing techniques. The algorithm for computing the beacon's position is detailed in Section~\ref{beacon_direction}. Section~\ref{Algorithm_Implementation} describes the implementation of the signal extraction and positioning algorithms. The validation of the proposed localization method through simulation is discussed in Section~\ref{sim_validation}. The system's implementation and corresponding laboratory experiments are presented in Section~\ref{lab_implementation}. Section~\ref{implementation} covers the integration of the system within the LARS and the AUV, along with the outcomes of pool experiments. Section~\ref{Discussion} analyzes the experimental results and provides suggestions for system enhancements. Finally, Section~\ref{Conclusions} summarizes the primary findings and presents concluding remarks.

\section{Physical Model and Mathematical Formulation}\label{Mathematical_Model}
\subsection{EM Beacon}

\begin{figure}[t]\centering
 \includegraphics[width=\linewidth]{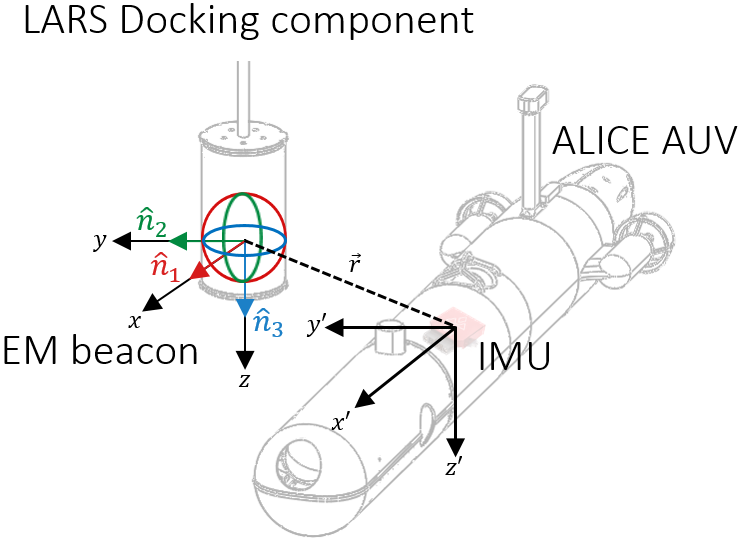}
  \caption{EM beacon and magnetometer coordinate frames.}
  \label{beacon_frame}
\end{figure}

The EM beacon used in this study consists of three transmitting coils, denoted $i=1,2,3$ and positioned in an orthogonal configuration as shown in Figure (\ref{beacon_frame}). To distinguish the beacon's artificial field from other magnetic fields, each coil is driven by a sinusoidal signal with a distinct frequency. Assuming the sensing magnetometer operates at distances exceeding the beacon's diameter but shorter than the signals' wavelength, we can model each coil as a magnetic dipole. In this work, the main coordinate frames are the EM beacon frame $[x,y,z]$ with its origin fixed to the beacon's center and the axes aligned with the coils and the AUV frame $[x', y', z']$, with its origin fixed to the AUV's magnetometer origin and aligned with the magnetometer (IMU) axes. Consequently, the magnetic flux density of the i'th coil can be expressed by \cite{cheng1989field}:

\begin{equation}{\label{biosavart}}
\begin{split}
\Vec{B}_{i}(x,y,z) = \frac{\mu}{4 \pi} \bigg[ \frac{3(\Vec M_{i} \cdot \Vec r) \Vec r - \Vec M_{i} r^2}{r^5} \bigg] 
\end{split}
\end{equation}

where the magnetic flux density at a given location is determined by the vector $\Vec{r}_{1x3} = [x , y, z]$, representing the distance from a specific location to the center of the EM beacon, located at $[x_0 , y_0, z_0]$. The permeability of the medium is denoted by $\mu = \mu_0 \cdot \mu_r$ with the permeability of the free space $\mu_0 = 4 \cdot \pi \cdot 10^{-7} [kg \cdot m \cdot s^{-2} \cdot A^{-2}]$ and the relative permeability of the medium $\mu_r$ (approximately 1 for water). The magnetic dipole moments of the three transmitting coils, denoted by $\Vec{M_i}$ are defined as follows:

\begin{equation}{\label{M}}
\Vec{M_i}= A_i \cdot N_i \cdot I_i \cdot sin (\omega_i t + \phi_i ) \cdot \hat{n}_i
\end{equation}
where $A_i$ represents the coil's cross-sectional area, $N_i$ denotes the number of turns, $I_i$ is the amplitude, $\omega_i$ is the frequency and $\phi_i$ is the phase of the excitation current of the coils. The orientation of each coil in the EM beacon frame is defined by $\hat{n_i}$, as shown in Figure (\ref{beacon_frame}) where:

\begin{equation}{\label{n}}
\begin{split}
\hat{n}_1 = [1 , 0, 0]^T  \qquad \hat{n}_2 = [0 , 1, 0]^T  \qquad \hat{n}_3 = [0 , 0, 1]^T 
\end{split}
\end{equation}

\subsection{Receiving Magnetometers}

The onboard magnetometer, a component of the AUV's Inertial Measurement Unit (IMU), measures the magnetic flux density along the AUV's body-fixed frame. The acquired signals encompass contributions from diverse sources, including the EM beacon signals, Earth's geomagnetic field, and ambient noise. We make the assumption that errors stemming from sensor cross-axis misalignment, nonlinearity, and scale factor can be disregarded. Consequently, we express the measured magnetic flux density, denoted as $\vec{B}'$, using the following model:
\begin{equation}\label{mag_field}
\vec{B}'^T(x,y,z) = \textbf{R} \cdot [\vec{B}^T(x,y,z) + \vec{B}_E^T(x,y,z) + noise] 
\end{equation}

where the combined magnetic flux density of the beacon's signals is represented by the vector $\vec{B} = \vec{B}_1 + \vec{B}_2 + \vec{B}_3$. The geomagnetic field, referenced to the beacon's body-fixed frame is denoted by $\vec{B}_E(x,y,z)$ and the transformation between the beacon's frame and the AUV's frame is provided by the matrix $\mathbf{R} = R_{\psi} R_{\vartheta} R_{\varphi}$, where $R_{\psi}$, $R_{\vartheta}$, and $R_{\varphi}$ represent rotations about the $x$, $y$ and $z$ axes, respectively, with:

\begin{equation}
R_{\varphi} =
\begin{pmatrix}
1     & 0           &  0          \\
0     & \cos \varphi   &  -\sin \varphi \\ 
0     & \sin \varphi   &   \cos \varphi 
\end{pmatrix}
\end{equation}

\begin{equation}
R_{\vartheta} =
\begin{pmatrix}
\cos \vartheta   & 0   &  \sin \vartheta  \\
0             & 1   &  0            \\ 
-\sin \vartheta  & 0   &   \cos \vartheta 
\end{pmatrix}
\end{equation}

\begin{equation}
R_{\psi} =
\begin{pmatrix}
\cos \psi   & -\sin \psi   &  0  \\
\sin \psi   & \cos \psi    &  0  \\ 
0           & 0            &  1 
\end{pmatrix}
\end{equation}

\section{Signal processing}\label{signal_processing}

The signal of each specific coil, denoted as $\vec{B}_{i}$ was extracted from the comprehensive magnetometer measurements using a digital implementation of lock-in amplifiers (LIA) \cite{zhang2020fpga}. These LIAs detect both the amplitudes and phases of signals that correlate with the internally-generated reference signals, as defined by: 

\begin{align} \label{ref_sig1}
s_{i} &= 2 \cdot sin (\omega_{i} t)  \\
s^{\frac{\pi}{2}}_{i} &= 2 \cdot cos (\omega_{i} t) 
\end{align}

where $\omega_{i}$ is the frequency of each coil and $s^{\frac{\pi}{2}}_{i}$ is a signal shifted by a phase of $\frac{\pi}{2}$ with respect to $s_{i}$. To identify the beacon signals, the sampled signal $\vec{B}'$ was multiplied by $s_{i}$ and $s^{\frac{\pi}{2}}_{i}$. For each LIA, signals other than the i'th signal were considered as noise. 

Denoting the combination of the geomagnetic field, other beacon signals and noise by $n(t)$, the outcome of this multiplication with the reference signals is expressed as follows:

\begin{equation}\label{multiply1}
\vec{B}' \cdot s_{i} = 2 \cdot |\textbf{R} \cdot \vec{B}_{i}| \cdot sin (\omega_{i} t + \phi_{i}) \cdot sin  (\omega_{i} t) + n(t)  \cdot  sin  (\omega_{i} t)
\end{equation}
\begin{equation}\label{multiply2}
\vec{B}' \cdot s^{\frac{\pi}{2}}_{i} = 2 \cdot |\textbf{R} \cdot \vec{B}_{i}| \cdot sin  (\omega_{i} t + \phi_{i}) \cdot cos (\omega_{i} t) + n(t)  \cdot  cos  (\omega_{i} t)
\end{equation}
Equations (\ref{multiply1})-(\ref{multiply2}) can be rewritten as:
\begin{equation} \label{identity1}
\vec{B}'  \cdot s_{i} =  |\textbf{R} \cdot \vec{B}_{i}| \cdot \big[ cos (\phi_{i}) - cos (2 \omega_{i} t + \phi_{i}) \big] + n(t)  \cdot  sin  (\omega_{i} t)
\end{equation}
\begin{equation} \label{identity2}
\vec{B}' \cdot s^{\frac{\pi}{2}}_{i} =   |\textbf{R} \cdot \vec{B}_{i}| \cdot \big[ sin ( 2 \omega_{i} t + \phi_{i}) + sin (\phi_{i}) \big]  + n(t)  \cdot  cos  (\omega_{i} t)
\end{equation}

Applying low-pass filters on Eq. \ref{identity1} and Eq. \ref{identity2} removes the time-dependent components in those while keeping the DC value of the signal:

\begin{equation} \label{lfp1}
\big[\vec{B}' \cdot s_{i} \big]_{LPF} =  |\textbf{R} \cdot \vec{B}_{i}| \cdot cos (\phi_{i})
\end{equation}
\begin{equation} \label{lfp2}
\big[\vec{B}' \cdot s^{\frac{\pi}{2}}_{i} \big]_{LPF} =   |\textbf{R} \cdot \vec{B}_{i}| \cdot sin (\phi_{i})
\end{equation}
Denoting the beacon's signal of the i'th coil, as measured by the magnetometer by $\vec{B'_i}$ and the phases by $\vec{\phi_{i}}$ (with respect to the reference signals), the magnitudes and the phases of the beacon's signals can be derived through:
\begin{equation} \label{amplitude}
|\vec{B'_i}| =  \sqrt{( \textbf{R} \cdot \vec{B}_{i} \cdot cos  (\phi_{i}))^2 + (\textbf{R} \cdot \vec{B}_{i} \cdot sin (\phi_{i}))^2}
\end{equation}

\begin{equation} \label{phase_lia}
\vec{\phi_{i}} =  tan^{-1}\bigg(\frac{|\textbf{R} \cdot \vec{B}_{i}| \cdot sin (\phi_{i})}{|\textbf{R} \cdot \vec{B}_{i}| \cdot cos  (\phi_{i})}\bigg)
\end{equation}

\begin{figure*}[t]
\centering
 \includegraphics[width=0.6\linewidth]{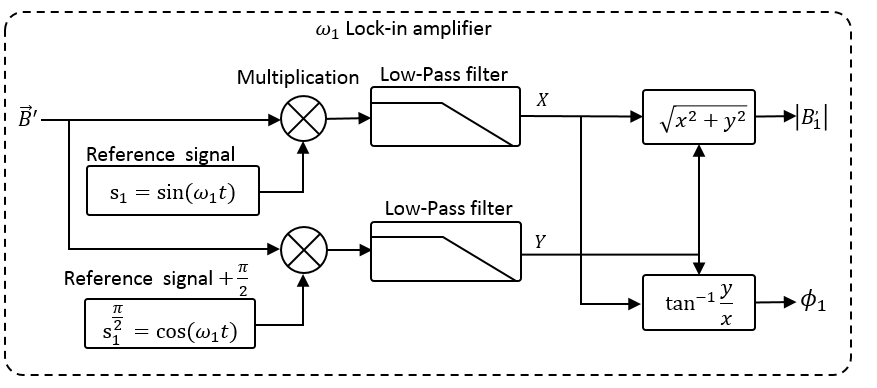}
  \caption{A schematic description of a lock-in amplifier for the extraction of a signal at a frequency $\omega_1$}
  \label{lock_in_amp}
\end{figure*}

A schematic description of the LIA components tuned for the extraction of the signal at the frequency $\omega_1$ is presented in Figure \ref{lock_in_amp}. 

\section{Computation of beacon location}\label{beacon_direction}

\subsection{Determination of beacon direction}
To determine the position $\Vec r$ and orientation $\textbf{R}$, the calculated magnitudes (Eq. (\ref{amplitude})) were incorporated into Eq.(\ref{biosavart}): 

\begin{equation}{\label{solv_eq}}
|\textbf{R}^T\vec{B}_{i}'| =\frac{\mu}{4 \pi} \bigg[ \frac{3(\Vec M_{i} \cdot \Vec r) \Vec r - \Vec M_{i} r^2}{r^5} \bigg] 
\end{equation}

\begin{figure*}[t]
         \centering
         \includegraphics[width=\textwidth]{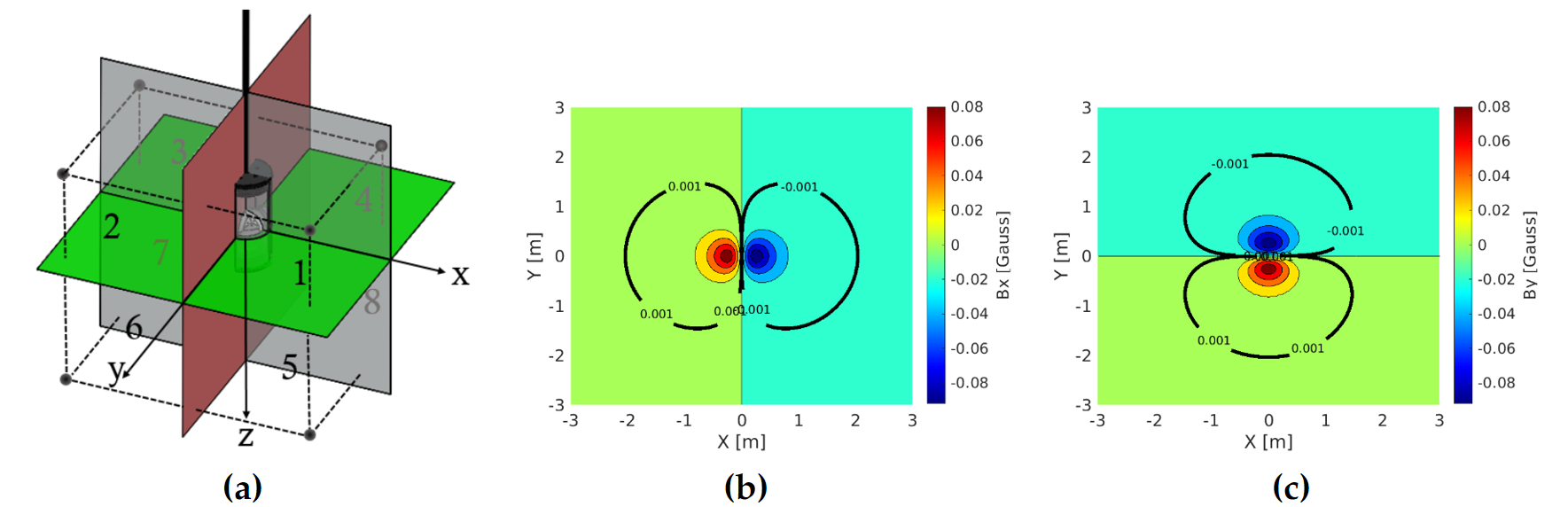}
         \caption{ (\textbf{a}) displays eight sectors of the beacon, highlighting points with identical magnetic field values. The magnetic flux density of ${B}^x_{1}$  (\textbf{b}) and ${B}^y_{1}$ on the plane $z = 0.5m$ was calculated by solving Eq. (\ref{biosavart}) for a coil with a diameter of 0.12 m, 370 turns, and a DC current of 3.2 A. Contours at the 0.001 Gauss level illustrate the dipole field's shape, with the $X=0$ line in (\textbf{b}) and $Y=0$ in  (\textbf{c}) indicating points where components of the magnetic flux density change direction.} 
          \label{field_sim} 
\end{figure*}

However, the electromagnetic field produced by the beacon exhibits multiple locations where the magnitudes of the magnetic flux density are equivalent in various regions, as illustrated in Figure \ref{field_sim}\textbf{a}. Consequently, determining a unique position is only possible after resolving the signs of $\vec{B_i}$. This challenge was overcome by analyzing the phases $\vec{\phi}_i$ computed by Eq.(\ref{phase_lia}) utilizing the property that, corresponding to the direction of the magnetic field, phases exhibit variations when intersecting the axes of the coil, as demonstrated in Figures \ref{field_sim}\textbf{b} and \ref{field_sim}\textbf{c}.

\begin{figure}[t]
\centering
 \includegraphics[width=\linewidth]{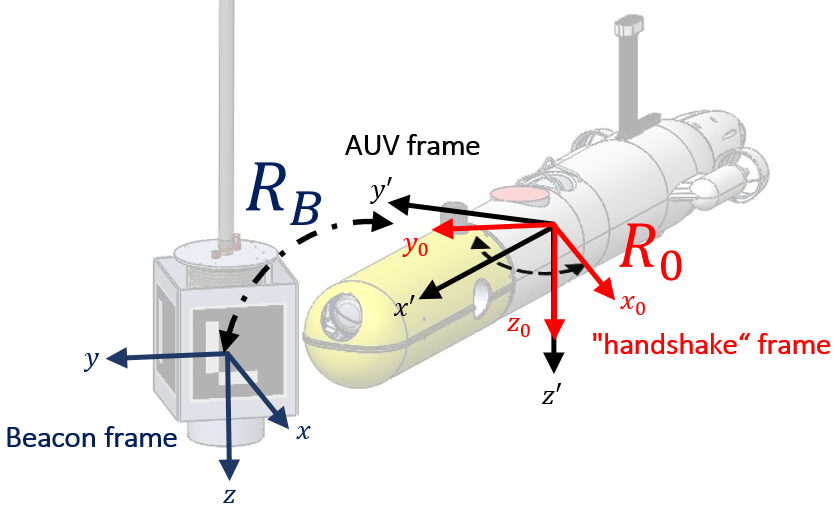}
  \caption{Spatial configuration of the AUV, Docking component, Beacon, Body and Reference frames at the "handshake" moment}
  \label{handshake}
\end{figure}

Since the EM beacon and reference signals operate in separate frames, an initial calibration process was employed to establish the spatial relationship between position coordinates and measured phases. This calibration occurred during a "handshake" between the vision and EM guidance phases, when both the AUV magnetometer and the forward-looking camera simultaneously detected the docking component, as depicted in Figure (\ref{handshake}). In this process, the relative position of the AUV with respect to the beacon was determined and a "handshake" frame was established. This frame, fixed to the AUV's origin and aligned with the beacon's frame at the "handshake" moment, allows the separation of the rotation of the beacon with respect to the AUV ($\textbf{R})$ into the rotation of the beacon ($\textbf{R}_B$) and the rotation of the AUV ($\textbf{R}_{0}$) with respect to the "handshake" frame, obtained from the AUV's navigation system. Consequently, the "handshake" frame enables the measurement of magnetic flux densities and phase variations independently of the AUV's orientation.

It is essential to note that phase computation is highly susceptible to interference. Therefore, achieving accurate positioning, which demands reliable signals from at least six field components, may not always be feasible. This challenge is particularly pronounced when the AUV is positioned near one of the beacon's axes or at a distance where signal weakening becomes noticeable.

To enhance the robustness of position computation, an alternative approach involves determining the relative direction to the beacon using either two components of one signal or three components of distinct signals. This is accomplished by assigning to each sector of the beacon a unique combination of in-phase and anti-phase components. However, as illustrated in Table (\ref{phase_pos}), each combination of phases corresponds to two potential solutions. For instance, sectors 1 and 7 share identical phase combinations, introducing ambiguity in identifying the correct solution. To address this ambiguity, the algorithm excludes transitions that are physically implausible or constrained by the dynamics of the AUV and the docking component. For example, it rules out direct transitions between sectors 1 and 7, as they would intersect with the docking component. Similarly, transitions between diagonal sectors 1 and 6 are considered implausible, given that the AUV performs decoupled vertical and horizontal movements during the terminal guidance phase.

 \begin{table}[h]
        \centering
	\begin{tabular}{|c|c|c|c|c|}
		\hline
	Sector &  Direction        &  ${\phi}_{1}^x$, ${\phi}_{1}^y$ &   ${\phi}_{2}^y$, ${\phi}_{2}^z$ &  ${\phi}_{3}^x$, ${\phi}_{3}^z$ \\\hline
        1      &  x > 0, y>0, z>0  &   + , +   & + , + & + , + \\\hline
        7      &  x < 0, y<0, z<0  &   + , +   & + , + & + , + \\\hline
 	2	   &  x > 0, y<0, z>0  &   + , -   & - , + & - , - \\\hline
   	8	   &  x < 0, y>0, z<0  &   + , -   & - , + & - , - \\\hline
 	3	   &  x < 0, y<0, z>0  &   - , -   & + , - & + , - \\\hline   
  	5	   &  x > 0, y>0, z<0  &   - , -   & + , - & + , - \\\hline 
        4	   &  x < 0, y>0, z>0  &   - , +   & - , - & - , + \\\hline		
 	6      &  x > 0, y<0, z<0  &   - , +   & - , - & - , + \\\hline
			\end{tabular}
	\caption{Relations between the signal phase and position, where "+" and "-" denotes the in-phase and anti-phase field components respectively. }
 \label{phase_pos}
\end{table}
\subsection{Computation of beacon position}

The position of the beacon was determined by solving the system of equations (\ref{solv_eq}), providing six variables representing position and orientation denoted $\Vec r$ and $\textbf{R}_B$. This system consists of nine equations, with each $\Vec{B}_i$ comprising three components ($B^x_i$, $B^y_i$, $B^z_i$). Consequently, the equations in (\ref{solv_eq}) form an over-determined system. To address this, an optimization technique, specifically the Levenberg-Marquardt (LM) method \cite{LM}, was employed to find values for $\Vec r$ and $\textbf{R}_B$ that minimize $\vec{f}_i$ as defined by:

\begin{equation}{\label{solv_lm}}
\vec{f}_i(\vec{r}, \textbf{R}_B) =\frac{\mu}{4 \pi} \bigg[ \frac{3(\Vec M_{i} \cdot \Vec r) \Vec r - \Vec M_{i} r^2}{r^5} \bigg] - |\textbf{R}_B \textbf{R}_{0}|^T \cdot \vec{B}'_i \cdot sign\{\vec{B}'_{i}\} 
\end{equation}
 
\section{Algorithm Implementation}\label{Algorithm_Implementation}

\begin{figure*}[t]
 \includegraphics[width=\linewidth]{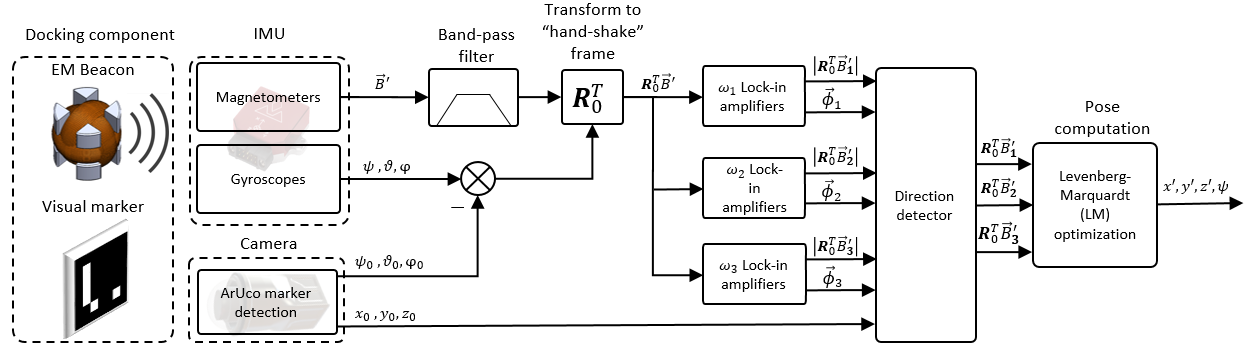}
  \caption{Schematic description of the signal extraction and the positioning algorithms}
  \label{schematic}
\end{figure*}

The developed signal extraction and positioning algorithms were implemented as ROS nodes \cite{quigley2009ros}. For signal processing filters, the real-time DSP IIR library was utilized and for position computation, the Mobile Robot Programming Toolkit (MRPT) library was employed. A schematic representation of the algorithms is presented in Figure (\ref{schematic}).

To mitigate the impact of the AUV movement on the measured signals and reduce the influence of high-frequency noise, the magnetometer signals underwent an initial filtration. This involved the use of a fourth-order Butterworth band-pass filter with a central frequency of 20 Hz and a bandwidth of 10 Hz. Following this filtration, the signals were transformed into the "handshake" frame, denoted as $\textbf{R}_0$. This frame was established by integrating the AUV's orientation, determined by the onboard navigation system and the beacon's orientation at the moment of the "handshake." The latter was identified using the ArUco marker detection algorithm \cite{kalaitzakis2020experimental}, which utilizes unique binary patterns and a simple detection algorithm to recognize each marker, computing its position and orientation.

Subsequently, the transformed signals were processed with lock-in amplifiers (LIA) to extract the beacon signals. This process involved multiplying the transformed signals by three time dependent sine and cosine functions corresponding to the beacon's frequencies. These products were then filtered using fourth-order Butterworth low-pass filters with a cutoff frequency of 0.4 Hz. The outputs from the filters underwent further processing to calculate $|\textbf{R}^T_0 \vec{B}'_i|$ and $\vec{\phi}'_i$ using Equations (\ref{amplitude}) and (\ref{phase_lia}). 

The obtained amplitudes and phases, along with the initial position $\vec{r_0}=[x_0,y_0,z_0]$, were inputted into a direction detection function to determine the signs of $\vec{B}'_i$ and the relative direction to the beacon.

This function utilized the initial position to set the signs of $\vec{B}'_i$ and employed a specialized zero-crossing detection logic. This logic involved switching the signs based on the measured phases, identifying instances where the phases shifted by $180^\circ$. To mitigate erroneous crossings caused by frequency discrepancies between the beacon and reference signals \cite{sonnaillon2007lock} or noise, the computed phases were continuously assessed using a moving window consisting of 200 measurements. Within this window, the periodic average of the signal, standard deviation and the rate of change of the phases were calculated. Empirically determined threshold values were applied to identify and exclude erroneous crossings.

To detect the relative position based only on phase information, a simplified version of Table \ref{phase_pos}, containing only four possible phase combinations (shown in Table \ref{phase_pos_reduced}), was utilized. This simplified table serves as a look-up table in conjunction with the motion constraints logic discussed in Section \ref{beacon_direction}. The specific procedure is elaborated in Algorithm \ref{state_alg}.

 \begin{table}[t]
        \centering
	\begin{tabular}{|c|c|c|c|c|}
		\hline
	Sector &  Direction        &  ${\phi}_{1}^x$, ${\phi}_{1}^y$ &   ${\phi}_{2}^y$, ${\phi}_{2}^z$ &  ${\phi}_{3}^x$, ${\phi}_{3}^z$ \\\hline
        1      &  x > 0, y>0, z>0  &   + , +   & + , + & + , + \\\hline
 	2	   &  x > 0, y<0, z>0  &   + , -   & - , + & - , - \\\hline
 	3	   &  x < 0, y<0, z>0  &   - , -   & + , - & + , - \\\hline   
        4	   &  x < 0, y>0, z>0  &   - , +   & - , - & - , + \\\hline 		
			\end{tabular}
	\caption{Reduced table of four possible combinations of phases, where "+" and "-" denotes the in-phase and anti-phase field components respectively. }
 \label{phase_pos_reduced}
\end{table}

\begin{algorithm}
  \caption{Direction computation algorithm}
  \label{state_alg}
  
\KwData{$\Vec{r}_0 , \textbf{R}_0, \vec{\phi}_1 , \vec{\phi}_2, \vec{\phi}_3$}
$\vec{\Phi} \gets \vec{\phi}_1 , \vec{\phi}_2, \vec{\phi}_3$\;
\KwResult{$sector$}

\If{ArUco marker detected}{
$ sector \gets \Vec r_0$\;
$Sector{\_}initialized$ = True}

\If{$Sector{\_}initialized$}{
$prev{\_}sector = sector$\;
$state$ = Look-up Table \ref{phase_pos_reduced} $(\vec{\Phi})$\;

            \uIf{$sector$ = 1 and $prev{\_}sector$ = 1 or 2 or 4 or 5}
                {$sector = 1$} 
            \uElseIf{$sector$ = 1 and $prev{\_}sector$ = 7 or 6 or 8 or 3}
                {$sector = 7$} 
            \uElseIf{$sector$ = 2 and $prev{\_}sector$ = 2 or 1 or 3 or 6}
                {$sector = 2$} 
            \uElseIf{$sector$ = 2 and $prev{\_}sector$ = 8 or 5 or 7 or 4}
                {$sector = 8$} 
            \uElseIf{$sector$ = 3 and $prev{\_}sector$ = 3 or 2 or 4 or 7}
                {$sector = 3$} 
            \uElseIf{$sector$ = 5 and $prev{\_}sector$ = 5 or 6 or 8 or 1}
                {$sector = 5$} 
            \uElseIf{$sector$ = 4 and $prev{\_}sector$ = 4 or 1 or 3 or 8}
                {$sector = 4$} 
            \uElseIf{$sector$ = 6 and $prev{\_}sector$ = 6 or 5 or 7 or 2}
                {$sector = 6$}
}      
\end{algorithm}

Ultimately, the AUV's position and orientation were determined using a real-time implementation of the Levenberg-Marquardt (LM) least squares method. This method resolves Equation (\ref{solv_lm}). To enhance the computation's robustness, the solver was constrained to provide a solution within the maximum detection range of 2.5 meters around the beacon. This constraint was realized by introducing the "penalty" function:
\begin{equation}
    f_4(\vec{r}) =
    \begin{cases}
        1 &  |\vec{r}| > 2.5 \\
        0 & \text{else }
    \end{cases}
\end{equation}

To enhance robustness, the number of computed variables was reduced by setting the values of the computed pitch and roll of the docking component to zero. This decision was based on the assumption that the docking component, fixed to the surface platform via a flexible wire, primarily induces heave motion due to surface waves, with significant pitch and roll motions not expected. Consequently, the angles $\vartheta$ and $\varphi$ were constrained to zero, reducing the number of independent variables and strengthening the robustness of the calculation. Furthermore, to prevent positioning errors resulting from weak signals, the position computation was activated only when at least one value of $B'$ exceeded a threshold of 0.03 Gauss.

To enhance the accuracy of the solution, a refinement process was implemented. This process involves detecting and excluding outliers that exceed the maximum expected positional variation based on the AUV's maximum speed during the terminal guidance maneuvers (0.15 m/s). Additionally, an averaging technique was applied over a moving window that includes 50 computed positions.

The complete description of the positioning algorithm is provided by pseudo Algorithm \ref{pose_alg}. 

\begin{algorithm}
  \caption{Position computation algorithm}
  \label{pose_alg}
    \KwData{$\vec{B}', \textbf{R}_0, \vec{r}_0$}
    \KwResult{$\vec{r}, \textbf{R}_B$}
    $prev{\_}\vec{\phi}_{1,2,3} \gets 0 $\;
    \If{$ArUco{\_}marker{\_}detected$}{
    $R_0 \gets  [\psi - \psi_0, \vartheta - \vartheta_0, \varphi - \varphi_0]$\;
    $\Vec{r}_0 \gets [x_0,y_0,z_0]$\;    
    \For{$i \gets 1$ to $3$}{
    $sign\{\Vec{B}'_{i}\} = \frac{\mu}{4 \pi} \bigg[ \frac{3(\Vec M_{i} \cdot \Vec r_0) \Vec r_0 - \Vec M_{i} r_0^2}{r_0^5} \bigg] \cdot \textbf{R}_0$;}    
    $Phase{\_}initialized$ = True}
    \If{$Phase{\_}initialized$}{
    \For{$i \gets 1$ to $3$}{
    $\big|\textbf{R}_0 \Vec{B}'_i\big|, \vec{\phi}_i \gets LIA_i\{\textbf{R}_0\Vec{B}'\}$ \; 
    \For{$j \gets 1$ to $3$}{
    \If{$\phi_i[j] \cdot prev{\_}\phi_i[j] < 0$}{
    $sign\{\Vec{B}'_{i}\} = - sign\{\Vec{B}'_{i}\}$}
    $prev{\_}\phi_i[j] = \phi_i[j]$
            }
        }
    $\vec{r}, \textbf{R}_B \gets \frac{\mu}{4 \pi} \bigg[ \frac{3(\Vec M_{i} \cdot \Vec r) \Vec r - \Vec M_{i} r^2}{r^5} \bigg] - \big|\textbf{R}_B^T \textbf{R}_{0}^T \cdot \vec{B}'_i \big| \cdot sign\{\vec{B}'_{i}\}$
    }  
\end{algorithm}

\section{Simulation-Based Validation}\label{sim_validation}

A dedicated beacon simulation was utilized to assess the performance of the developed algorithms in signal extraction and positioning. This simulation generated a synthetic magnetic flux density using Equation (\ref{mag_field}), considering specific beacon properties outlined in Table \ref{beacon_prop}, along with parameters such as relative position $\vec{r}$ and orientation $\textbf{R}_B$.

To mimic the magnetometer properties described in Table \ref{sensor_prop} and accurately simulate real-world measurements, the signals were discretized based on the magnetometer's sampling rate and resolution. White noise with a standard deviation of 2 $mGauss$ was incorporated, and a geomagnetic field vector of $\vec{B}_E=[0.2, 0.13, 0.35]$ $Gauss$ was defined.

\begin{table}[t]
        \centering
	\begin{tabular}{ |l|l| }
		\hline	
	Beacon core diameter        & 120  mm \\
        Wire cross section area & 0.7 mm \\
        Coil 1 input current    & 1.53 A \\
        Coil 2 input current    & 1.3 A \\
        Coil 3 input current    & 1.4 A \\
        Coil 1 frequency        & 16 Hz \\
        Coil 2 frequency        & 20 Hz \\
        Coil 3 frequency        & 25 Hz \\
        Coil 1 number of turns  & 370 \\
        Coil 2 number of turns  & 370 \\
        Coil 3 number of turns  & 370 \\ 
		\hline
			\end{tabular}
	\caption{Transmitting beacon properties}
 \label{beacon_prop}

        \centering
	\begin{tabular}{ |l|l| }\hline
            Magnetometer  &  \\ 
  		\hline
		Model              & Vectornav VN100/VN300 \\
            Type           & MEMS \\
		Scale              & ±2.5 Gauss      \\
            Noise Density  & 140 $\mu$ Gauss/ $\sqrt{Hz}$  \\
		Resolution         & 1.5 mGauss      \\
		Sampling frequency & 200 Hz      \\
		\hline
			\end{tabular}
	\caption{Magnetometer properties}
 \label{sensor_prop}
\end{table}

To comprehensively assess the algorithm's robustness against potential modeling errors and sensor misalignment encountered in real-world applications, the simulation was repeated with simulated sensor misalignment errors. These errors included angular deviations of $5^o$ in roll, pitch, and yaw between the magnetometer and the AUV along with a 10\% error in modeling magnetic moments $M_{i}$.

The algorithms underwent evaluation in two distinct scenarios. The first focused on static positioning, introducing 16 fixed positions evenly spaced at intervals of 10 cm. Positioning accuracy was assessed using the root mean square error (RMSE) value calculated over 600 solutions for each point.

\begin{figure}[t]
  \centering
    \includegraphics[width=\linewidth]{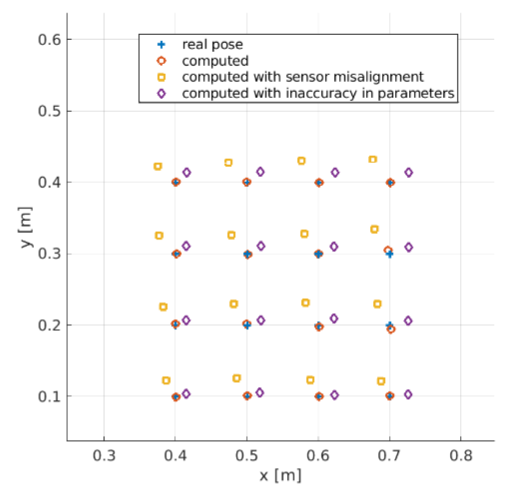}
    \caption{Simulation results for position computation in selected points.} \label{sim_results_static}
\end{figure}

The results of the static positioning simulation, as illustrated in Figure \ref{sim_results_static}, indicated an accuracy of 3 mm with precisely configured beacon parameters, 2.5 cm in the presence of some uncertainty in the beacon parameters and 5.2 cm when simulating magnetometer misalignment.

\begin{figure}[t]
    \centering
    \includegraphics[width=\linewidth]{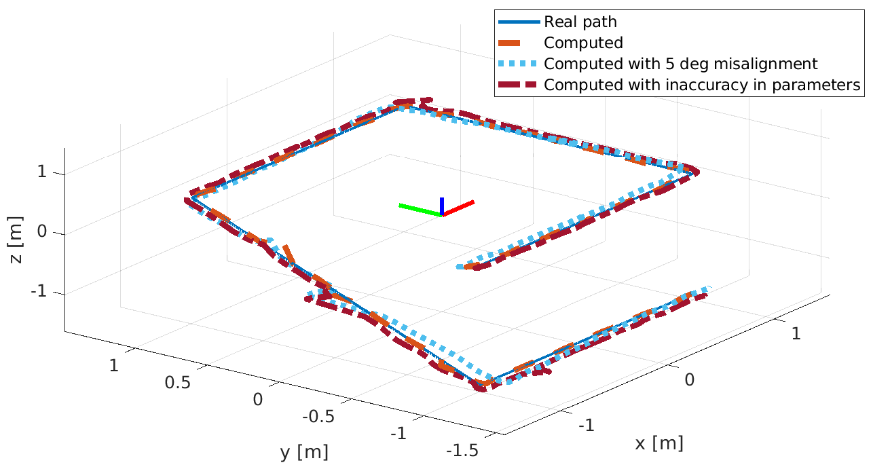}
    \caption{Simulation results for position computation along a path.} \label{sim_results_dynamic}
  \end{figure}

In the dynamic scenario, linear motion was simulated at a constant speed of 0.1 m/s along a predefined path. In this scenario, as presented in Figure \ref{sim_results_dynamic}, an accuracy of 13 cm was achieved with accurate knowledge of the beacon parameters, 15 cm in the presence of uncertainties in the beacon parameters and 14 cm with applied magnetometer misalignment. A summary of the results for static and dynamic simulations is provided in Table \ref{sim_results_table}. These findings confirm the algorithm's capability to compute positions with sufficient accuracy for guiding the AUV to the docking position. However, it is evident that precision levels are notably affected by onboard magnetometer misalignment and inaccuracies in the beacon's properties. To enhance position accuracy, these parameters should be verified and carefully adjusted.

\begin{table}[t]
        \centering
	\begin{tabular}{ |l|l| }
		\hline	
	Scenario                                     &  RMSE [m]\\
        Static                                   & 0.0031 \\
        Static with inaccuracy in parameters     & 0.0252 \\
        Static with sensor misalignment          & 0.052 \\
        Dynamic                                  & 0.13 \\
        Dynamic with inaccuracy in parameters    & 0.148 \\
        Dynamic with sensor misalignment         & 0.144 \\

		\hline
			\end{tabular}
	\caption{Summery of the simulation results}
 \label{sim_results_table}
\end{table}

\section{In-Laboratory Prototype Testing}\label{lab_implementation}

\subsection{Lab System implementation}
\begin{figure}[t]
\centering
 \includegraphics[width=\linewidth]{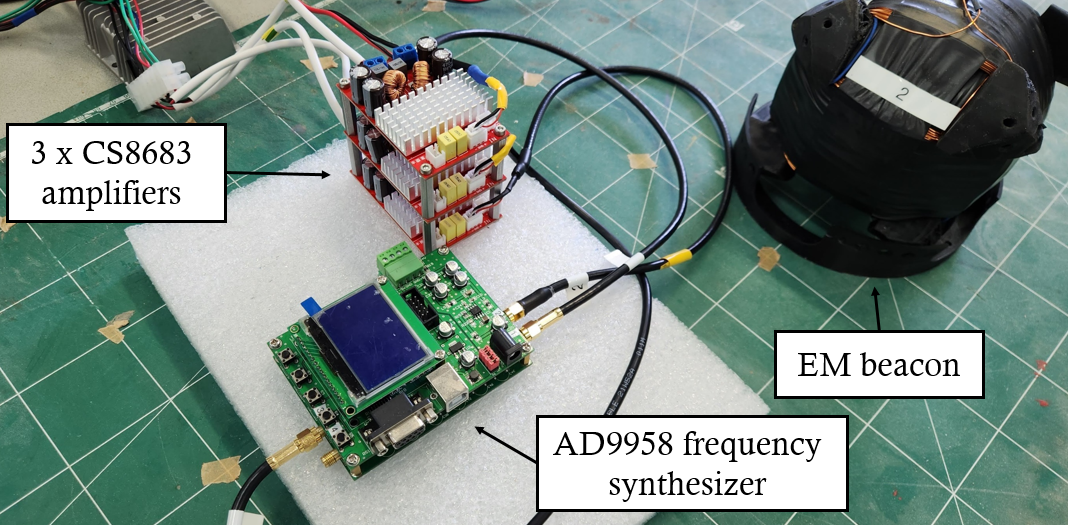}
  \caption{Prototype of the experimental EM beacon and the signal generation system in Lab experiment}
  \label{beacon}
\end{figure}

A functional prototype of the beacon and the signal generation system was developed for real-world testing, as illustrated in Figure (\ref{beacon}). The construction of the beacon involved winding three orthogonal copper coils around a 3D-printed ABS core with a 12-centimeter diameter. Each coil comprised approximately 370 turns. The beacon's dimensions were designed to fit into the 6-inch cylindrical housing of the docking component. The coils were powered by three alternating sine signals, generated by an AD9958 multi-channel frequency synthesizer and subsequently amplified using three mono-channel 130W, 24 VDC CS8683 digital amplifiers. These signals were configured to operate at frequencies of 16 Hz, 20 Hz, and 25 Hz to ensure adequate sampling by the receiving magnetometer.

In the laboratory experiments, the magnetic flux density was measured using a VectorNav VN-300 Inertial Navigation System (INS) equipped with a 200 Hz tri-axial magnetometer, gyroscopes, and accelerometers. This system was linked to a PC for sampling the measurements and executing the signal extraction and positioning algorithms outlined in Section \ref{Algorithm_Implementation}. Comprehensive specifications for both the beacon and the magnetometer were previously provided in Tables (\ref{beacon_prop}) and (\ref{sensor_prop}), respectively.

In order to achieve a precise calibration of the input signals, the amplified signals were initially measured using an oscilloscope. The measurements indicated frequency errors of $~0.6\%$, as summarized in Table \ref{scope_measure}. However, while these errors were minimal, they may introducing a significant measurement errors and effect the stability of the output signal \cite{sonnaillon2007lock}. 

\begin{figure}[t]
\centering
 \includegraphics[width=\linewidth]{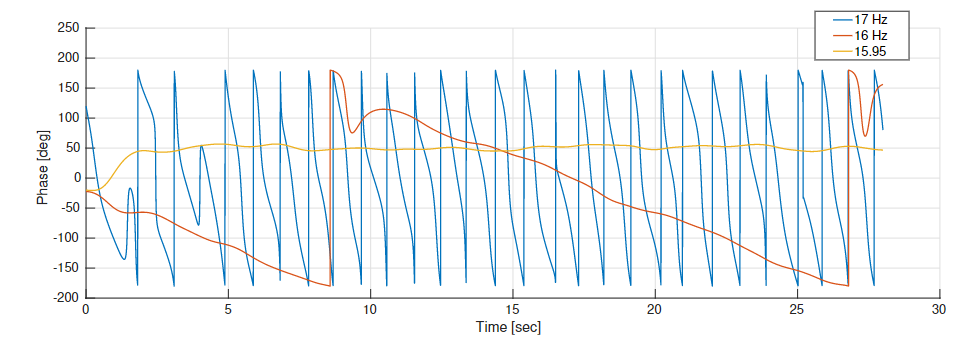}
  \caption{Phase computation from a single field component as computed for three different frequencies of reference signals.}
  \label{phase_drift}
\end{figure}

To mitigate these issues, fine-tuning of the reference signals was conducted by adjusting their frequencies until the output of the phase exhibited stable and consistent values. Figure \ref{phase_drift} illustrates this process, demonstrating phase computation inconsistency during static measurement, even with a minor error of 0.3\% in the reference signal frequency. The figure also illustrates the successful tuning of the signal achieved at a frequency of 15.95 Hz.

\begin{table}[t]
        \centering
	\begin{tabular}{ |l|l| }\hline
	Preset frequency & Measured frequency (\% max error) \\ \hline
	16 $Hz$ & 15.91 - 15.96  $Hz$ (0.56 \%)\\
	20 $Hz$ & 19.86 - 19.95  $Hz$ (0.7 \%) \\
	25 $Hz$ & 24.85 - 24.98  $Hz$ (0.6 \%)\\
		\hline
			\end{tabular}
	\caption{Summery of the frequency errors as measured by the oscilloscope}
 \label{scope_measure}
\end{table}

In addition, cross-talk voltages \cite{bian2021induced} resulting from interactions between the beacon coils were measured and their impact was assessed. To measure cross-talk voltage, a signal was supplied to one coil while simultaneously the induced voltage was measured on the other coils. The recorded cross-talk voltages consistently stayed below 30 mV during the test, confirming their negligible influence on the overall beacon signal.

\subsection{Lab System implementation}

\begin{figure}[t]
\centering
 \includegraphics[width=\linewidth]{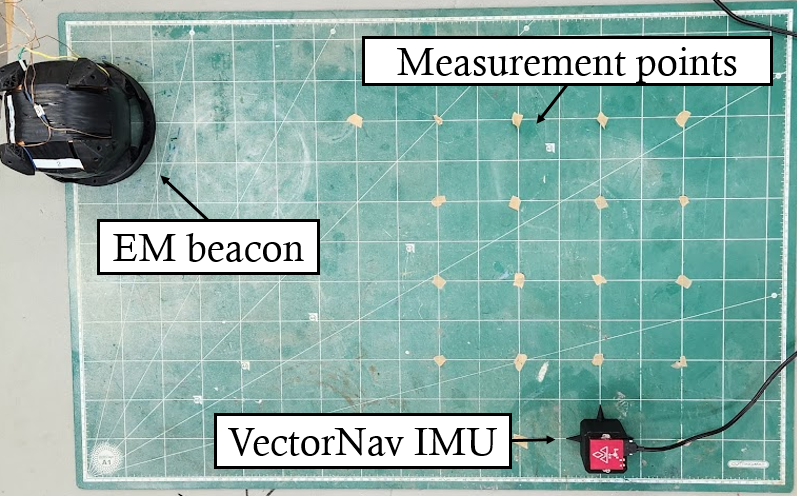}
  \caption{Setup of the lab experiment}
  \label{lab_test}
\end{figure}

The system's positioning accuracy was assessed in a controlled laboratory experiment. In this setup, the beacon was placed on a flat plane with grid intervals of 5 cm. Much like the static simulation scenario, the magnetometer was positioned at 16 discrete points in close proximity to the beacon, enabling a direct comparison between the algorithm's results and the known positions. This experimental configuration is illustrated in Figure (\ref{lab_test}).

To initialize the algorithm, the position data from the first measurement point was used. The precision of the computed positions was carefully evaluated by measuring the error between the average value of 45 positioning solutions and the actual position of each point. To account for installation errors, each point was adjusted by an overall offset vector $\vec{X}_{offset} = [0.102, 0.058, 0.003]$. This offset was based on the average error between the computed and the actual points, computed by: 

\begin{equation}
\vec{X}_{offset} = \frac{1}{16}\sum_{i = 1..16}( \vec{X}^p_i  - \langle \vec{X}^{c}_i \rangle ) 
\end{equation}

Here, $\vec{X}^p$ represents the actual position and $\langle \vec{X}^{c}_i \rangle$ denotes the averaged position derived from 45 solutions for each point. This adjustment procedure aimed to mitigate errors related to the installation of the beacon. Additionally, an assessment was conducted by measuring the relative distance between every pair of adjacent points, leveraging the inherent precision of the measuring grid.

\begin{figure}[t]
\centering
 \includegraphics[width=\linewidth]{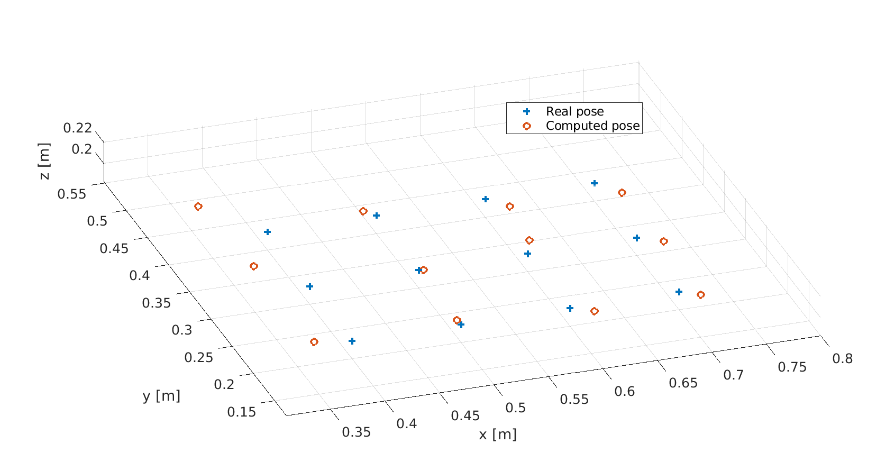}
  \caption{Positioning results for 16 measurement points in the lab experiment.}
  \label{points_exp}
\end{figure}

\begin{figure}[t]
    \includegraphics[width=\linewidth]{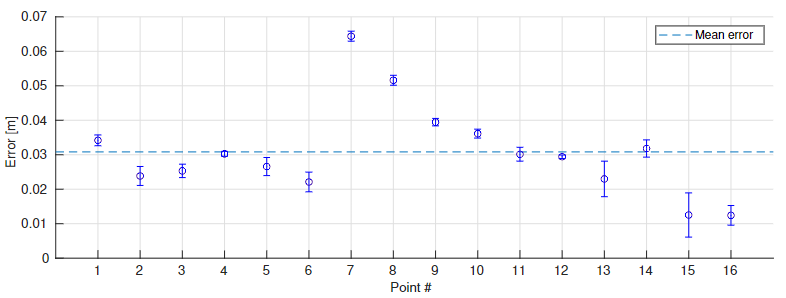}
    \caption{Positioning errors with respect to the measurement points in the Lab experiment.} \label{points_error_results}
  \end{figure}

\begin{figure}[t]
    \includegraphics[width=\linewidth]{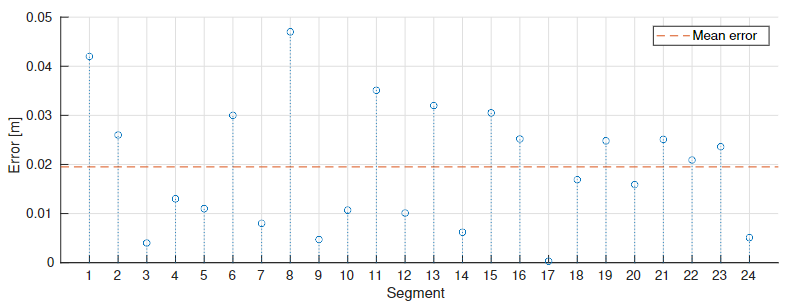}
\caption{Relative distance errors with respect to the measurement grid segments in the Lab experiment.}
\label{segment_error_results}
\end{figure}

The experimental outcomes, depicted in Figure (\ref{points_exp}), along with the positioning errors illustrated in Figures (\ref{points_error_results})- (\ref{segment_error_results}), highlighted the system's real-time capability to calculate the position relative to the beacon, achieving an average accuracy of under 3 cm. However, it's crucial to acknowledge that inaccuracies stemming from variations in the coils' properties and the installation process can significantly affect positioning precision. Hence, a systematic calibration procedure is essential before operational usage to rectify such discrepancies.

\section{System integration and experiments}\label{implementation}

\begin{figure*}[t]
\centering
 \includegraphics[width=\linewidth]{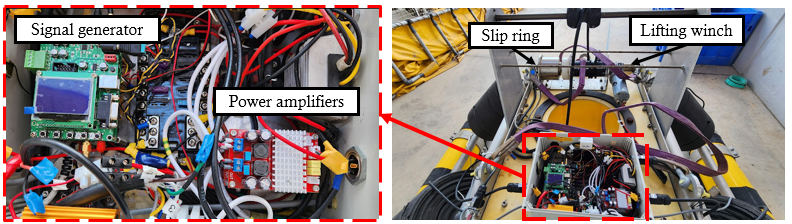}
  \caption{The signal generator and the power amplifiers inside the electronics box of the LARS system}
  \label{lars}
\end{figure*}

Following the successful outcomes of the laboratory experiments, the developed system was integrated into the LARS. The signal generator and amplifiers were installed within an electronic enclosure located on the LARS floating platform, as depicted in Figure (\ref{lars}). A remote power-enable switching circuit was implemented to enable remote activation of the beacon when the platform was deployed. The amplified signals were transferred to the beacon via a slip ring mechanism, enabling the deployment of the docking component to depths of up to 10 meters.

\begin{figure}[t]
\centering
 \includegraphics[width=0.8\linewidth]{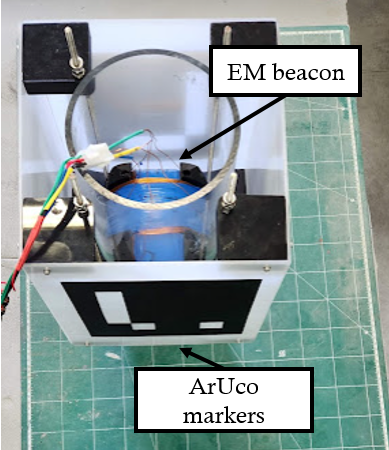}
  \caption{A prototype of the experimental EM beacon and signal generation system in Lab experiment}
  \label{housing}
\end{figure}

The beacon was integrated into the docking component. For heat management, the beacon housing was filled with 3M NOVEC™ 7100 engineered fluid to improve heat dissipation from the beacon. Furthermore, four ArUco markers were positioned around the housing to facilitate the visual guidance phase and the "handshake" process. These markers were arranged in a configuration that allowed capturing the beacon's position and orientation irrespective of the approach direction, enabling omnidirectional docking. The installation of the beacon inside the housing and the positioning of the visual markers are illustrated in Figure (\ref{housing}).

Regarding the AUV's sensor setup, the initial selection was the AUV's native magnetometer, integrated with the ADIS16488A INS. However, due to its relatively modest sampling rate of 123 Hz, the VectorNav VN-100 INS, equipped with a magnetometer similar to the VN-300, was integrated to provide a higher sampling rate. Furthermore, the AUV's onboard camera (Allied Vision Prosilica GT6600) was employed to detect the visual markers. In addition, the AUV's navigation filter was incorporated into the positioning algorithm to provide continuous tracking.

\begin{figure*}[t]
\centering
 \includegraphics[width=0.9\linewidth]{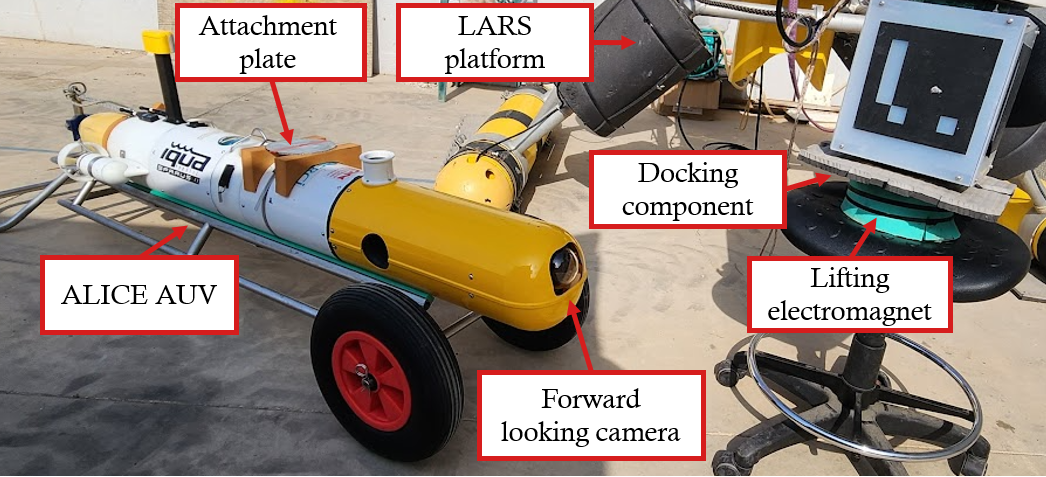}
  \caption{The attachment plate located on ALICE and the lifting electromagnet embedded at the bottom of the docking component.}
  \label{plate}
\end{figure*}

To latch the AUV, a lifting electromagnet was installed beneath the housing of the docking component and a metal attachment plate was fixed on top the AUV's hull near its center of gravity, as shown in Figure (\ref{plate}).

\subsection{Assessment of accuracy and detection range in the integrated System}

\begin{figure}[t]
\centering
    \includegraphics[width=\linewidth]{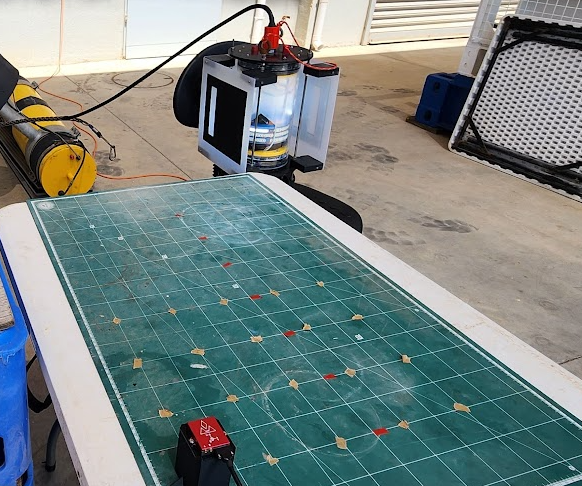}
    \caption{The docking component of the LARS, with the EM-beacon and the ArUco markers in an experiment.} \label{dock_outside}
\end{figure}

\begin{figure}[t]
\centering
    \includegraphics[width=\linewidth]{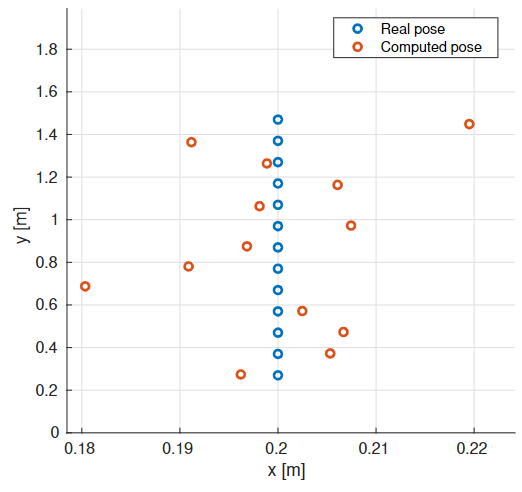}
    \caption{Positioning results of the integrated system in a controlled experiment} \label{distance_points}
  \end{figure}

\begin{figure}[t]
  \centering
    \includegraphics[width=\linewidth]{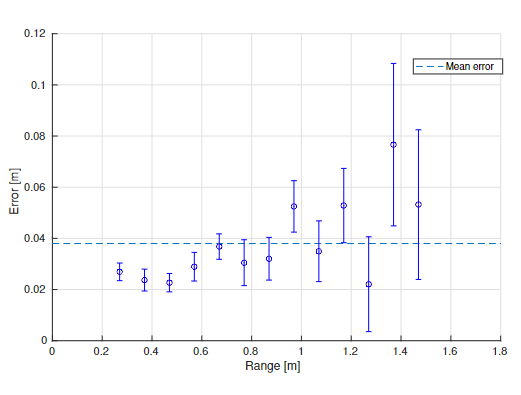}
    \caption{Error evaluation for the integrated system in a controlled experiment} \label{error_distance_points}
\end{figure}

Before conducting docking experiments in the pool, the integrated system underwent assessment in an experiment focused on detection distance and accuracy. To test the detection range of the system, the Vectornav VN300 INS was placed at 13 equidistant points, spaced at 10 cm intervals up to 1.5 meters, as shown in Figure (\ref{dock_outside}). The positioning results, depicted in Figure (\ref{distance_points}) with the error analyses, presented in Figure (\ref{error_distance_points}) demonstrated that the system successfully detected the beacon at distances up to 1.5 m with sufficient accuracy of less than 4 cm up to the range of 0.9 m and 10 cm up to 1.5 m. Notably, the average errors and standard deviations increased beyond a distance of 1 m. However, for the specific task of docking, where the highest precision is required as the AUV approaches very close to the docking component, the system achieved the necessary level of accuracy.

\subsection{Docking maneuvering sequence}

The maneuvering sequence during the visual and EM guidance phases utilized the ALICE AUV capability for decoupled motion control in heave, sway, surge and yaw \cite{gutnik2022adaptation}.  In the visual guidance phase, the control system was activated to position the AUV in the "handshake" position, maintaining a distance of 1.4 meters between its center of gravity and the beacon while facing towards the marker. This specific distance was chosen to position the onboard magnetometer as close as possible to the beacon while ensuring continued visual detection of the markers. Furthermore, after completing this phase, magnetometer measurements were excluded from AUV's navigation filter. This exclusion was necessary in close proximity to the docking component, where measurements of the geomagnetic field are significantly disturbed by the magnetic fields of the beacon and the electromagnetic lifting device.

Following the initialization of the EM positioning, the terminal docking sequence was activated. To ensure precise docking and avoid potential collisions between the docking component and AUV's hull and appendages, the terminal docking sequence employed a two-phase motion control strategy. 

\begin{figure}[h]
\centering
 \includegraphics[width=0.8\linewidth]{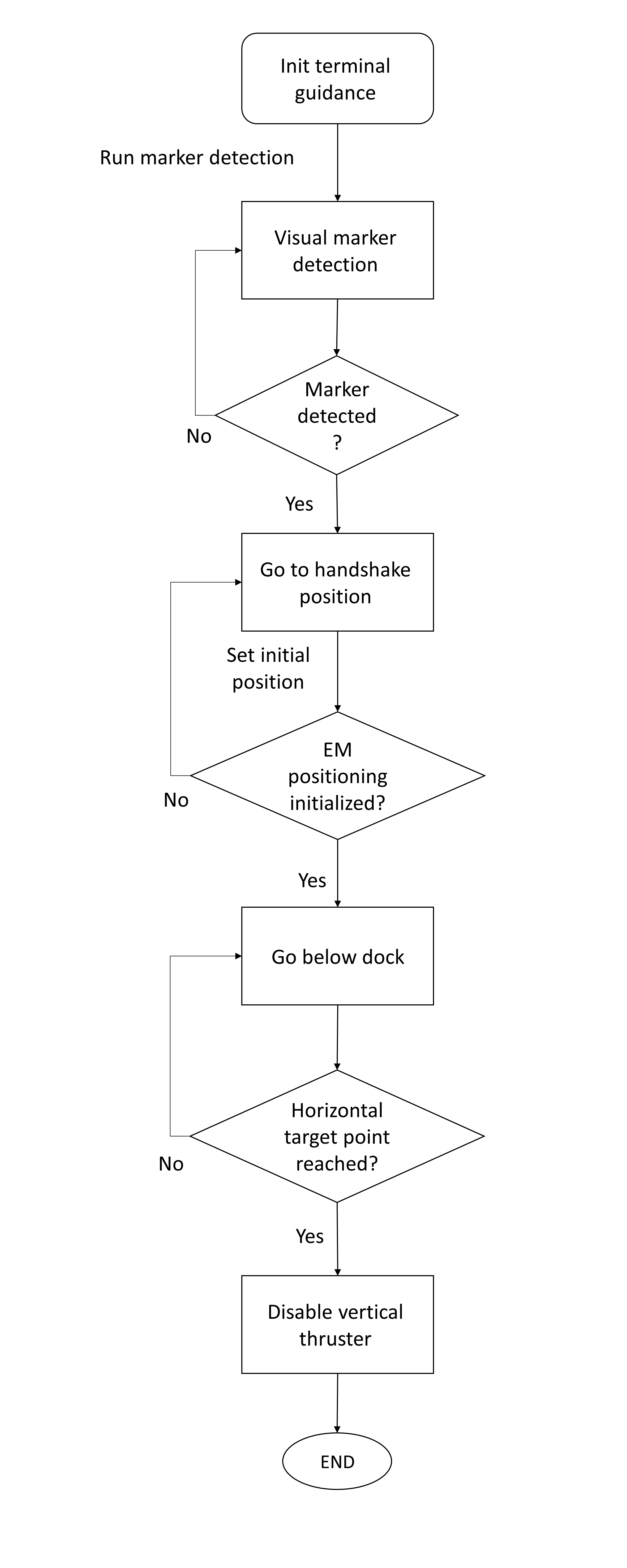}
  \caption{Schematic description of the terminal guidance phase}
  \label{sequence}
\end{figure}

In the first phase of the terminal guidance maneuver, AUV descended to a depth of 0.5 meters below the docking component. Once it reached this depth, the surge and sway PID controllers were activated to horizontally position AUV beneath the docking component with a margin of 15 cm. Within this range, the vertical thruster was turned off, allowing AUV's positive buoyancy to gradually lift it until it securely attached to the lifting electromagnet. Throughout the terminal guidance, the AUV's surge and sway motion controllers were limited to a maximum speed of 0.1 m/s. The complete docking process is outlined in Figure (\ref{sequence}).

\subsection{Docking experiments in a pool}

Ultimately, the system's ability to provide guidance during the terminal phase of docking was assessed in a complete real pool environment. The test occurred in our 9x3x2.8 (L X W X H) meter sea water pool, involving both the entire configuration of the ALICE AUV and the docking component.

\begin{figure}[!h]
  \centering
    \includegraphics[width=\linewidth]{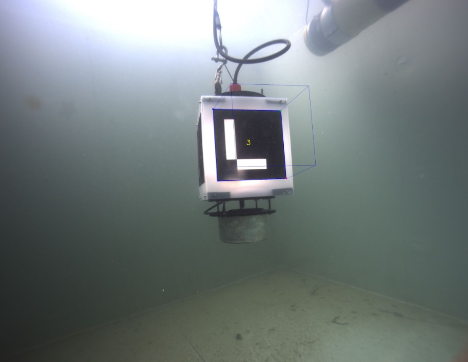}
    \caption{The docking component in the pool experiment, as identified by ALICE's onboard camera and the ArUco marker detection algorithm.} \label{pool_aruco}
  \end{figure}

\begin{figure}[!h]
  \centering
 \includegraphics[width=\linewidth]{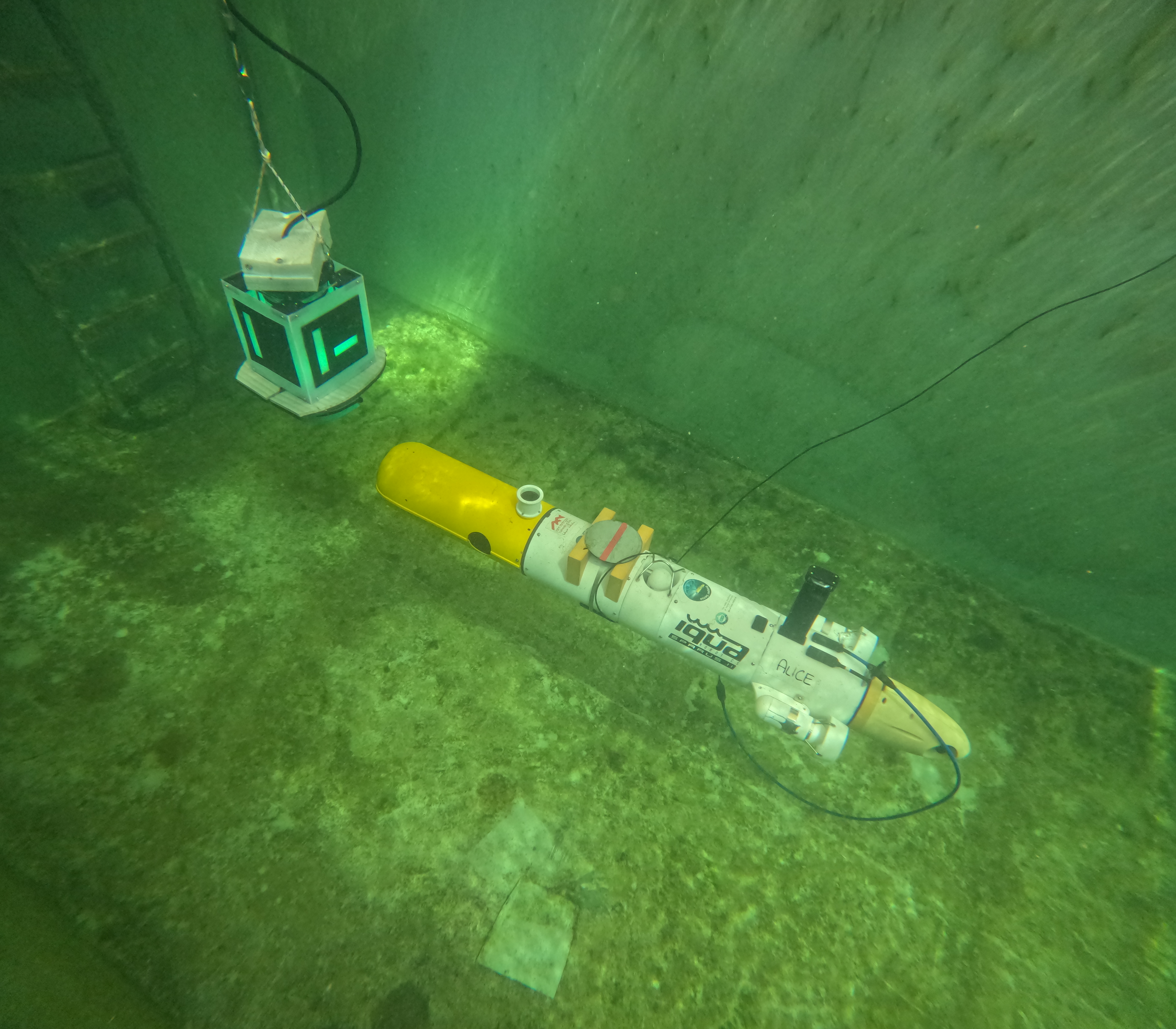}
    \caption{The setup of the pool experiment, consisting the omnidirectional docking component and the ALICE AUV} \label{pool_exp_1}
  \end{figure}

In this experiment, the docking component was submerged to a depth of approximately 1.5 meters below the surface, with the AUV positioned at a distance where the specified marker detection algorithm could detect the ArUco markers. At this point, the docking sequence commenced, and the AUV autonomously approached the "handshake" position using the markers, as depicted in Figure (\ref{pool_aruco}). When the AUV reached a distance of 1.4 meters from the beacon, the position of the beacon, determined from the markers, was used to initiate the EM guidance. After initialization, the AUV descended to a depth of 0.5 meters below the docking component, as depicted in Figure (\ref{pool_exp_1}). At this phase, the AUV maintained a consistent depth and regulated its lateral and forward movements based on the EM positioning algorithm until reaching the predetermined tolerance of 15 cm below the component.


\section{Discussion} \label{Discussion}

The experimental results demonstrated the system's capability to achieve accurate positioning solutions up to 1.5 meters. However, during the docking maneuver in the pool experiment, the AUV sometimes moved beyond this range, resulting in a loss of positioning. To extend the effective range of the EM guidance, a potential solution is to provide the beacon with a higher magnetic power.

Additionally, the sampling rate of the VectorNav IMU determines the beacon's maximum frequency for proper signal sampling. However, lower frequencies require low-pass filters within the LIAs. Consequently, filters with lower cutoff frequencies increase response times, introducing delays in the signal extraction algorithm. Thus, magnetometers with higher sampling rates could significantly enhance the positioning solution. Lastly, enhancing the positioning algorithm using state-of-the-art machine learning and deep learning methods, as suggested in Gutnik et al.'s work \cite{gutnik2023data}, might provide a robust computation approach tailored for this specific application.

\section{Conclusions} \label{Conclusions}

In this research, a novel method for 3D EM guidance was introduced to facilitate the precise, omnidirectional navigation of an AUV towards a docking component of a LARS. Key contributions of this study include the design of a compact EM beacon and its signal generation system, along with the creation of dedicated algorithms for real-time signal extraction and 3D positioning, compatible with commonly available magnetometers. To enable comprehensive 3D positioning, a novel initialization process aided by a vision-based positioning algorithm was incorporated. A significant contribution of this method, when compared with others, is its capability to perform precise docking where a direct line of sight between the LARS component and the AUV is not required. The system's development and the algorithms' performance were rigorously evaluated using dedicated simulations. Following this, a series of experiments were conducted to assess the system's accuracy and effectiveness in guiding our ALICE AUV during the final docking phase. The experimental results demonstrated high accuracy, achieving less than 4 cm within a range of 0.8 m and 10 cm within a range of 1.5 m. These results highlight the method's capability to effectively guide the AUV during the docking process. In the final phase of the study, the EM beacon system was integrated into our LARS, and the algorithmic framework was implemented in the ALICE AUV control system to guide it through an actual docking task in a controlled pool environment. The experiments conducted in this controlled setting validated the system's ability to provide precise positioning guidance during the critical terminal docking phase.

\ifCLASSOPTIONcaptionsoff
  \newpage
\fi




\bibliographystyle{IEEEtran}


\begin{thebibliography}{10}
\providecommand{\url}[1]{#1}
\csname url@samestyle\endcsname
\providecommand{\newblock}{\relax}
\providecommand{\bibinfo}[2]{#2}
\providecommand{\BIBentrySTDinterwordspacing}{\spaceskip=0pt\relax}
\providecommand{\BIBentryALTinterwordstretchfactor}{4}
\providecommand{\BIBentryALTinterwordspacing}{\spaceskip=\fontdimen2\font plus
\BIBentryALTinterwordstretchfactor\fontdimen3\font minus \fontdimen4\font\relax}
\providecommand{\BIBforeignlanguage}[2]{{%
\expandafter\ifx\csname l@#1\endcsname\relax
\typeout{** WARNING: IEEEtran.bst: No hyphenation pattern has been}%
\typeout{** loaded for the language `#1'. Using the pattern for}%
\typeout{** the default language instead.}%
\else
\language=\csname l@#1\endcsname
\fi
#2}}
\providecommand{\BIBdecl}{\relax}
\BIBdecl
\renewcommand{\BIBentryALTinterwordstretchfactor}{4}

\bibitem{gu2018automated}
H.-t. Gu, L.~Meng, G.~Bai, H.~Zhang, Y.~Lin, and S.~Liu, ``Automated recovery of the uuv based on the towed system by the usv,'' in \emph{2018 OCEANS-MTS/IEEE Kobe Techno-Oceans (OTO)}.\hskip 1em plus 0.5em minus 0.4em\relax IEEE, 2018, pp. 1--7.

\bibitem{sarda2016usv}
E.~I. Sarda and M.~R. Dhanak, ``A usv-based automated launch and recovery system for auvs,'' \emph{IEEE journal of oceanic engineering}, vol.~42, no.~1, pp. 37--55, 2016.

\bibitem{jalving2008payload}
B.~Jalving, J.~E. Faugstadmo, K.~Vestgard, O.~Hegrenaes, O.~Engelhardtsen, and B.~Hyland, ``Payload sensors, navigation and risk reduction for auv under ice surveys,'' in \emph{OCEANS 2008}.\hskip 1em plus 0.5em minus 0.4em\relax IEEE, 2008, pp. 1--8.

\bibitem{lin2022docking}
M.~Lin, R.~Lin, C.~Yang, D.~Li, Z.~Zhang, Y.~Zhao, and W.~Ding, ``Docking to an underwater suspended charging station: Systematic design and experimental tests,'' \emph{Ocean Engineering}, vol. 249, p. 110766, 2022.

\bibitem{kimball2018artemis}
P.~W. Kimball, E.~B. Clark, M.~Scully, K.~Richmond, C.~Flesher, L.~E. Lindzey, J.~Harman, K.~Huffstutler, J.~Lawrence, S.~Lelievre \emph{et~al.}, ``The artemis under-ice auv docking system,'' \emph{Journal of field robotics}, vol.~35, no.~2, pp. 299--308, 2018.

\bibitem{sarda2018launch}
E.~I. Sarda and M.~R. Dhanak, ``Launch and recovery of an autonomous underwater vehicle from a station-keeping unmanned surface vehicle,'' \emph{IEEE Journal of Oceanic Engineering}, vol.~44, no.~2, pp. 290--299, 2018.

\bibitem{piskura2016development}
J.~C. Piskura, M.~Purcell, R.~Stokey, T.~Austin, D.~Tebo, R.~Christensen, and F.~Jaffre, ``Development of a robust line capture, line recovery (lclr) technology for autonomous docking of auvs,'' in \emph{OCEANS 2016 MTS/IEEE Monterey}.\hskip 1em plus 0.5em minus 0.4em\relax IEEE, 2016, pp. 1--5.

\bibitem{hildebrandt2017combining}
M.~Hildebrandt, L.~Christensen, and F.~Kirchner, ``Combining cameras, magnetometers and machine-learning into a close-range localization system for docking and homing,'' in \emph{OCEANS 2017-Anchorage}.\hskip 1em plus 0.5em minus 0.4em\relax IEEE, 2017, pp. 1--6.

\bibitem{miranda2013homing}
M.~Miranda, P.-P. Beaujean, E.~An, and M.~Dhanak, ``Homing an unmanned underwater vehicle equipped with a dusbl to an unmanned surface platform: A feasibility study,'' in \emph{2013 OCEANS-San Diego}.\hskip 1em plus 0.5em minus 0.4em\relax IEEE, 2013, pp. 1--10.

\bibitem{maki2013docking}
T.~Maki, R.~Shiroku, Y.~Sato, T.~Matsuda, T.~Sakamaki, and T.~Ura, ``Docking method for hovering type auvs by acoustic and visual positioning,'' in \emph{2013 IEEE international underwater technology symposium (UT)}.\hskip 1em plus 0.5em minus 0.4em\relax IEEE, 2013, pp. 1--6.

\bibitem{fan2019auv}
S.~Fan, C.~Liu, B.~Li, Y.~Xu, and W.~Xu, ``Auv docking based on usbl navigation and vision guidance,'' \emph{Journal of Marine Science and Technology}, vol.~24, no.~3, pp. 673--685, 2019.

\bibitem{kusche2021indoor}
R.~Kusche, S.~O. Schmidt, and H.~Hellbr{\"u}ck, ``Indoor positioning via artificial magnetic fields,'' \emph{IEEE Transactions on Instrumentation and Measurement}, vol.~70, pp. 1--9, 2021.

\bibitem{dai20176}
H.~Dai, S.~Song, X.~Zeng, S.~Su, M.~Lin, and M.~Q.-H. Meng, ``6-d electromagnetic tracking approach using uniaxial transmitting coil and tri-axial magneto-resistive sensor,'' \emph{IEEE Sensors Journal}, vol.~18, no.~3, pp. 1178--1186, 2017.

\bibitem{pasku2017magnetic}
V.~Pasku, A.~De~Angelis, G.~De~Angelis, D.~D. Arumugam, M.~Dionigi, P.~Carbone, A.~Moschitta, and D.~S. Ricketts, ``Magnetic field-based positioning systems,'' \emph{IEEE Communications Surveys \& Tutorials}, vol.~19, no.~3, pp. 2003--2017, 2017.

\bibitem{gutnik2023data}
Y.~Gutnik, N.~Cohen, I.~Klein, and M.~Groper, ``Data-driven underwater navigation workshop: Auv close-range localization and guidance employing an electro-magnetic beacon,'' in \emph{2023 IEEE Underwater Technology (UT)}.\hskip 1em plus 0.5em minus 0.4em\relax IEEE, 2023, pp. 1--5.

\bibitem{bian2021induced}
S.~Bian, P.~Hevesi, L.~Christensen, and P.~Lukowicz, ``Induced magnetic field-based indoor positioning system for underwater environments,'' \emph{Sensors}, vol.~21, no.~6, p. 2218, 2021.

\bibitem{sheinker2013localization}
A.~Sheinker, B.~Ginzburg, N.~Salomonski, L.~Frumkis, and B.-Z. Kaplan, ``Localization in 3-d using beacons of low frequency magnetic field,'' \emph{IEEE transactions on instrumentation and measurement}, vol.~62, no.~12, pp. 3194--3201, 2013.

\bibitem{andria2019development}
G.~Andria, F.~Attivissimo, A.~Di~Nisio, A.~M.~L. Lanzolla, P.~Larizza, and S.~Selicato, ``Development and performance evaluation of an electromagnetic tracking system for surgery navigation,'' \emph{Measurement}, vol. 148, p. 106916, 2019.

\bibitem{hu2012novel}
C.~Hu, S.~Song, X.~Wang, M.~Q.-H. Meng, and B.~Li, ``A novel positioning and orientation system based on three-axis magnetic coils,'' \emph{IEEE Transactions on Magnetics}, vol.~48, no.~7, pp. 2211--2219, 2012.

\bibitem{feezor2001autonomous}
M.~D. Feezor, F.~Y. Sorrell, P.~R. Blankinship, and J.~G. Bellingham, ``Autonomous underwater vehicle homing/docking via electromagnetic guidance,'' \emph{IEEE Journal of Oceanic Engineering}, vol.~26, no.~4, pp. 515--521, 2001.

\bibitem{peng2019low}
S.~Peng, J.~Liu, J.~Wu, C.~Li, B.~Liu, W.~Cai, and H.~Yu, ``A low-cost electromagnetic docking guidance system for micro autonomous underwater vehicles,'' \emph{Sensors}, vol.~19, no.~3, p. 682, 2019.

\bibitem{vandavasi2018concept}
B.~N.~J. Vandavasi, U.~Arunachalam, V.~Narayanaswamy, R.~Raju, D.~P. Vittal, R.~Muthiah, and A.~R. Gidugu, ``Concept and testing of an electromagnetic homing guidance system for autonomous underwater vehicles,'' \emph{Applied Ocean Research}, vol.~73, pp. 149--159, 2018.

\bibitem{lin2022underwater}
R.~Lin, Y.~Zhao, D.~Li, M.~Lin, and C.~Yang, ``Underwater electromagnetic guidance based on the magnetic dipole model applied in auv terminal docking,'' \emph{Journal of Marine Science and Engineering}, vol.~10, no.~7, p. 995, 2022.

\bibitem{cheng1989field}
D.~K. Cheng \emph{et~al.}, \emph{Field and wave electromagnetics}.\hskip 1em plus 0.5em minus 0.4em\relax Pearson Education India, 1989.

\bibitem{zhang2020fpga}
C.~Zhang, H.~Liu, J.~Ge, and H.~Dong, ``Fpga-based digital lock-in amplifier with high-precision automatic frequency tracking,'' \emph{Ieee Access}, vol.~8, pp. 123\,114--123\,122, 2020.

\bibitem{LM}
J.~J. Mor{\'e}, ``The levenberg-marquardt algorithm: Implementation and theory,'' in \emph{Numerical Analysis}, G.~A. Watson, Ed.\hskip 1em plus 0.5em minus 0.4em\relax Berlin, Heidelberg: Springer Berlin Heidelberg, 1978, pp. 105--116.

\bibitem{quigley2009ros}
M.~Quigley, K.~Conley, B.~Gerkey, J.~Faust, T.~Foote, J.~Leibs, R.~Wheeler, and A.~Y. Ng, ``\textsc{ROS: AN OPEN-SOURCE ROBOT OPERATING SYSTEM},'' in \emph{Proc. ICRA workshop on open source software}, vol.~3, no. 3.2, (Kobe, Japan), 2009, p.~5.

\bibitem{kalaitzakis2020experimental}
M.~Kalaitzakis, S.~Carroll, A.~Ambrosi, C.~Whitehead, and N.~Vitzilaios, ``Experimental comparison of fiducial markers for pose estimation,'' in \emph{2020 International Conference on Unmanned Aircraft Systems (ICUAS)}.\hskip 1em plus 0.5em minus 0.4em\relax IEEE, 2020, pp. 781--789.

\bibitem{sonnaillon2007lock}
M.~O. Sonnaillon and F.~J. Bonetto, ``Lock-in amplifier error prediction and correction in frequency sweep measurements,'' \emph{Review of scientific instruments}, vol.~78, no.~1, 2007.

\bibitem{gutnik2022adaptation}
Y.~Gutnik, A.~Avni, T.~Treibitz, and M.~Groper, ``On the adaptation of an auv into a dedicated platform for close range imaging survey missions,'' \emph{Journal of Marine Science and Engineering}, vol.~10, no.~7, p. 974, 2022.

\end{thebibliography}


\vfill


\end{document}